# Shapiro Steps in the Absence of Microwave Radiation

Lin He[1,2*], Jian Wang[2,3*], Moses H. W. Chan[2]

[1] Department of Physics, Beijing Normal University, Beijing, 100875, People's Republic of China
[2] The Center for Nanoscale Science and Department of Physics, The Pennsylvania State University, University Park, Pennsylvania 16802-6300, USA
[3] International Center for Quantum Materials and State Key Laboratory for Mesoscopic Physics, School of Physics, Peking University, Beijing, 100871, People's Republic of China.

The current-voltage *I-V* characteristics of a 1.2 μm long Au nanowire contacted by superconducting electrodes were studied in details. Interestingly, the *I-V* curves over a wide range of temperatures display multiple steps at voltages $V = (m/n)(V_0/2e)$ in the absence of microwave radiation, where $m$, $n$ are integer numbers and $V_0 \sim 92$ μV. We posit that these steps are the subharmonic Shapiro steps due to the interplay of the ac Josephson current and a minigap in the Au nanowire induced by superconducting electrodes via the proximity effect.

Although the proximity effect where superconductivity is induced in a normal metal due to its close proximity to a superconductor is a well studied subject, new phenomena are still be uncovered [1-3]. In a recent experiment, the transport properties of a Au nanowire of 70 nm in diameter contacted by four superconducting W electrodes were reported [4]. The distance between the two inner edges of voltage electrodes, *i.e.*, the length of the Au wire, is 1.2 μm. This system shows that the proximity induced superconductivity of the wire is acquired in two steps and it is characterized by a minigap $\delta$ instead of the superconducting gap $\Delta$ of the W electrode. Evidence for the minigap, in which the local density of states for superconducting quasiparticles is exactly zero [5-7] over a range of energy $[-\delta, +\delta]$ around the Fermi energy, in proximity structures was also revealed in recent scanning tunnelling microscopy studies [8-10] and another transport measurements [11]. The ratio of $\delta/\Delta$ decreases with increasing length of the normal metal between the superconducting electrodes [5-10].

In this Letter, the current-voltage *I-V* characteristics of the 1.2 μm long Au nanowire reported in reference 4 were studied in details. Eleven steps at specific voltages $V = (m/n)V_0/2e$ are clearly observed in the *I-V* curves over a wide range of temperatures, where $m$, $n$ are integers and $V_0 = 92$ μV. We attribute these steps to the Shapiro steps arising from the interplay between ac Josephson current and the minigap in the Au nanowire.

The mechanism that explains the physics of Josephson junctions is Andreev reflection where an electron incident from the normal metal is converted into a hole moving in the opposite direction [12,13], thus creating simultaneously a Cooper pair in superconductor. In a superconductor-normal metal nanowire-superconductor (S-NW-S) junction subjected to a low voltage bias $V$, the transport is dominated by multiple Andreev reflections (MARs) [14-23]. Due to MARs, there is time dependent alternating current in the Josephson junction, which can be written as a Fourier series $I(t) = \sum_n I_n e^{in\omega t}$ with the Josephson frequency $\omega = 2eV/\hbar$. Under a microwave radiation, the Josephson junction shows both harmonic and subharmonic Shapiro steps at voltages $V = (m/n)\hbar\omega_r/2e$, where $m$, $n$ are integers and $\omega_r$ is the radiation frequency [24-31]. Shapiro steps appear when the Josephson and microwave frequencies are commensurate ($n\omega = m\omega_r$) [30]. We will explain below a possible mechanism for the appearance of Shapiro steps in our experiment in the absence of microwave radiation.

In our experiment, the W strips used as the electrodes are composed of tungsten, carbon, and gallium and show a superconducting transition temperature of $T_C \sim 5.1$ K [32,33] and a critical magnetic field $H_C$ of $\sim 70$ kOe for temperature $T < 2.4$ K. Temperature dependent resistance *R-T* measurements showed that the Au wire acquires superconductivity via the proximity effect from the W electrodes in two steps and becomes completely superconducting below 3.5 K at zero magnetic field. Magneto-resistance measurements below 2.4 K show zero resistance up to a field of 40 kOe. Measurements between 2.6 and 3.4 K, however, show zero resistance at low magnetic field but an abrupt increase in $R$ of $\sim 8$ Ω at $H_V =$ 2.5 kOe. The resistance of the wire stays flat 8 Ω over a wide range of magnetic field above 2.5 kOe and increases to the normal state ($\sim 155$ Ω) only when the field is increased close to the critical field of the W electrode ($H_C \sim 50$ kOe) at $T \sim 2.6$ K. This W-Au NW-W junction shows characteristic zero resistance temperature and magnetic field of 3.5 K and 2.5 kOe that are completely different from that of the W electrodes. The proximity induced superconductivity in this junction is hence characterized by a minigap, $\delta$ [4]. The value of $\delta$ was not determined in reference 4. Interestingly it can be deduced from the Shapiro steps results to be presented below.

Fig. 1(a) shows the resistance of the W-Au NW-W junction as a function of the dc excitation or bias current *i.e.*, $R(I)$ curves, measured at various temperatures (the resistance is calculated by $R = V/I$). At 1.8 K, the critical current is easily identified by the sharp jump of the resistance. By increasing the temperature to above 2.0 K, a footlike structure at low resistance is seen between $I_{C1}$ and $I_{C2}$ in the $R(I)$ curves. Below $I_{C1}$, the voltage (or the resistance) generated in the junction is zero. Above $I_{C3}$, the junction is in the normal state. Similar footlike structure has been previously observed in superconducting microbridges attributing to nonequilibrium processes in the presence of a gap oscillation [34,35], and has also been observed in superconductor-normal metal-superconductor junction configurations arising from MARs [29,36,37]. In a S-NW-S junction, as in our system, the minigap in the normal metal is independent of the position [7,10]. This eliminates the nonequilibrium effect in the presence of the gap oscillation as the origin of the footlike structures. Thus the footlike structures in our W-Au NW-W junction is likely a consequence of MARs.

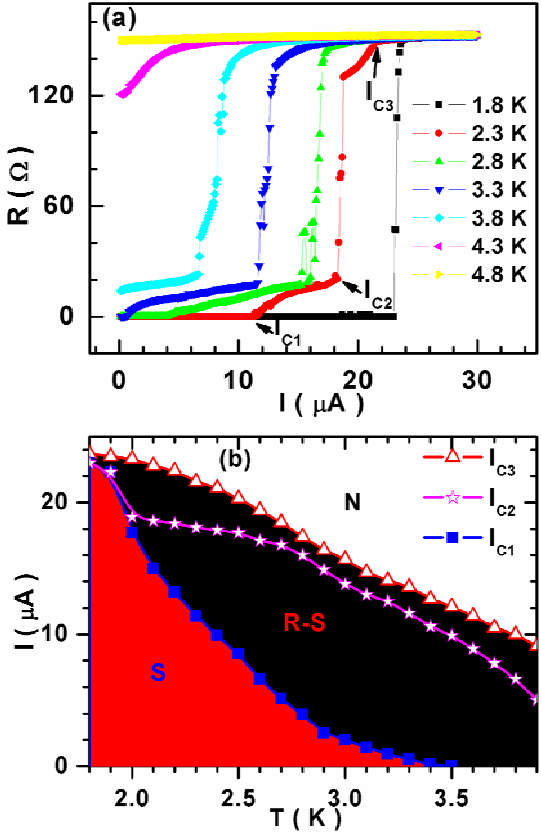

**FIG. 1.** (a) The resistance of the nanowire as a function of current $R(I)$ measured at various temperatures. $I_{C1}$ is the current below which the measured resistance of the junction is zero. $I_{C3}$ is the current above which the junction is in the normal state. Between 2 and 3.9 K, there is an intermediate critical current $I_{C2}$. A footlike structure in the $R(I)$ curves is seen between $I_{C1}$ and $I_{C2}$. (b) The $I$ vs $T$ phase diagram of the Au wire in the W-Au NW-W junction. Below $I_{C1}$, the wire is in superconductor state (S). Above $I_{C3}$, the wire is in normal state (N). The wire is in the resistance-superconducting state (R-S) when measured with a current between $I_{C1}$ and $I_{C3}$.

Figure 1(b) shows the phase diagram of the W-Au NW-W junction in the $I$-$T$ plane. Below $I_{C1}$, the junction is in the superconducting (S) state showing zero resistance. The temperature dependence of $I_{C1}$ was shown in Fig. 4(b) of Ref. 4. The value of $I_{C1}$ decreases with temperature. At 3.5 K, $I_{C1} = 0$, in good agreement with the magnetic field-temperature phase diagram shown in Fig. 4(a) of Ref. 4.

Above $I_{C3}$, the junction is in the normal (N) state and the resistance of the wire is about 155 Ω. Between $I_{C1}$ and $I_{C3}$, the junction is in the resistance-superconducting (R-S) state or finite-voltage state. The footlike structures between $I_{C1}$ and $I_{C2}$ are observed above 2.0 K, which indicates that MARs processes are important above 2 K. We will show below that the minigap plays an important role in the transport properties of the W-Au NW-W junction between $I_{C1}$ and $I_{C2}$.

We now turn to the main subject of the present Letter—the steps in the footlike structures between $I_{C1}$ and $I_{C2}$ in the absence of microwave radiation. Fig. 2(a) shows current-voltage characteristics of the W-Au NW-W junction at temperatures between 2.6 and 3.2 K. Several steps in the curves are clearly observed at specific voltages. Interestingly these steps can be labeled at $V = (m/n)V_0/2e$, where $m$, $n$ are integers and $V_0 = 92$ μV (index 1 in the figure). We noted that $V_0 = 92$ μV is a "fitting parameter" in that with such a choice, all the steps in the $I$ vs $V$ curves can be accounted for by the relation $V = (m/n)V_0/2e$. The voltage steps are more clearly visible by differentiating the $I$ vs $V$ curves of panel (a) numerically by a three points spline fit procedure. Clear peaks or dips are observed at voltages $V = (m/n)V_0/2e$ in these "d$I$/d$V$" curves as shown in Fig. 2(b). The positions of these peaks or dips appear to be independent of the temperatures between 2.1 and 3.7 K. The observed features are suggestive of the harmonic and subharmonic Shapiro steps at voltages $V = (m/n)\hbar\omega_r/2e$ observed under microwave radiations [24-31]. Figure 2(c) summarizes the peak and dip positions in the "d$I$/d$V$" curves of the W-Au NW-W junction as a function of temperature. The peaks and dips occur more frequently at integer steps ($m/n = 1$ and 2). This is similar to that observed in Ref. 29. Interestingly, these steps in our experiment are observed in the absence of microwave radiation. We noted that Hoffmann, *et al.* also observed a large number of peaks in the differential conductance d$I$/d$V$ curves without microwave radiations in superconductor-normal metal-superconductor junctions of different lengths [38]. Their peaks appear to be similar to that shown in Fig. 2(b). The authors of Ref. 38 however did not comment on the origin of the peaks appearing between 60 and 100 μV.

As should be noted, microwave radiation from a 'standard' external source is not always required for the observation of Shapiro steps. In the classic experiment of Giaever [39], a voltage-biased junction was used as a microwave source to induce Shapiro steps in a Josephson junction placed a few micrometers away [1,39]. However, the Shapiro steps can only be observed just slightly below $T_C$ (from 0.99$T_C$ to about 0.9$T_C$). Recently, it has been predicted theoretically that Shapiro steps can be generated by coupling a tunnel Josephson junction to a mechanical oscillator [40] or to a superconductor-normal metal-superconductor junction [41]. Obviously, the steps observed in our system have a different origin. The superconducting gap 2Δ = 1.54 meV of the W electrodes can be deduced from the superconducting transition temperature $T_C$ = 5.1 K. This is much larger than $eV_0$ = 92

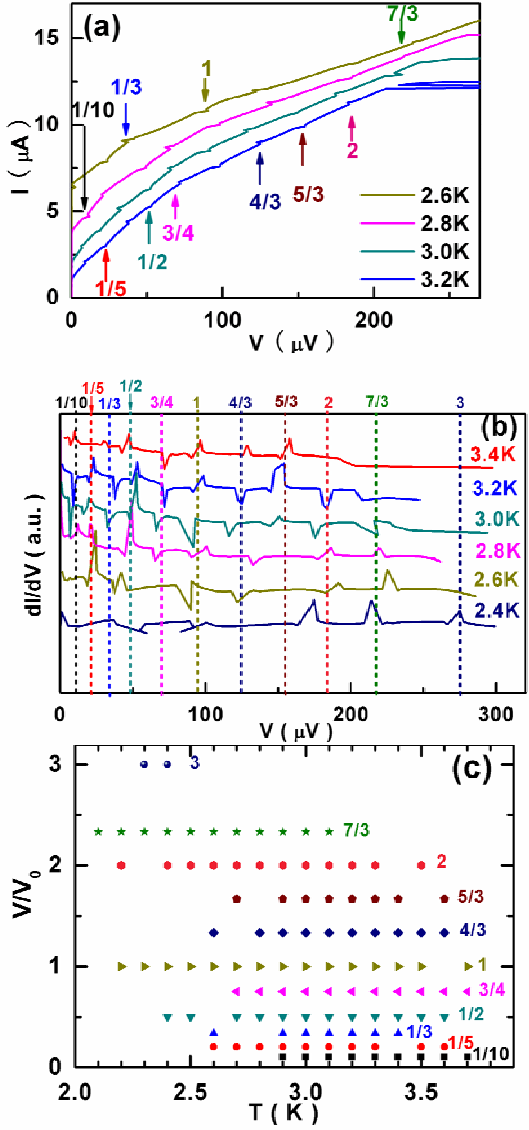

**FIG. 2.** (a) Typical current voltage curves measured at 2.6 K, 2.8 K, 3.0 K, and 3.2 K. Some of the voltage steps in the curves are identified by arrows. (b) $dI/dV$ vs $V$ obtained by numerically differentiating the curves of panel (a). Peaks and dips are clearly found at voltages about $V = (m/n)V_0/2e$ in the absence of microwave radiation. Here, $V_0 = 92$ µV (index 1 in the figure) and $m$, $n$ are integers. (c) Peak and dip positions in the "$dI/dV$" curves of the nanowire as a function of temperatures. The Y-axis is plotted in units of $V_0$.

µeV and excludes the W electrodes as the origin of the voltage steps in the current-voltage curves.

What is the origin of the voltage steps at voltages $V = (m/n)V_0/2e$ in the current-voltage curves? We attribute these steps to the Shapiro steps arising from the interplay between ac Josephson current and the minigap. Recently, Fuechsle, *et al.* investigated the effect of microwaves on the current-phase relation of long superconductor-normal metal-superconductor (Nb-Ag-Nb) Josephson junction and observed that the current-phase relation is strongly affected by microwave radiation [42]. This effect was attributed to the microwave excitation of quasiparticles in low-lying Andreev bound states (ABS) across the minigap in the normal metal to ABS carrying supercurrent in the opposite direction. In our W-Au NW-W Josephson junction, there are multiple ABS with energy that lie between the minigap of the NW, $\delta$, and the superconducting gap $\Delta$ of the electrode with a sharp peak near $\delta$ as illustrated in Fig. 3. In the presence of an excitation (to be discussed below) of appropriate energy, the quasiparticles in low-lying ABS will be excited across the minigap of the Au nanowire in the W-Au NW-W junction. The quantum jump from the ABS above $+\delta$ to the ABS below $-\delta$ will emit a microwave with frequency of $2\delta/\hbar$. As a consequence, the interplay between the ac Josephson current and the emitted microwave leads to Shapiro steps at voltages $V = (m/n)(2\delta/2e)$ when their frequencies are commensurate ($n\omega = m2\delta/\hbar$). In this model $2\delta$ is equivalent to $V_0$ as described above and a Shapiro step at $V = (m/n)(2\delta/2e)$ is visible only when the corresponding ac Josephson current of the Fourier series, $I_n$, in the junction gives a significant contribution [30]. Below we will give a justification of this explanation according to our experimental results and discuss the candidate of the excitation in our W-Au NW-W junction.

In S-NW-S junctions, the ratio of $\delta/\Delta$ decreases with increasing ratio of $L/\xi$ [7,10], where $L$ is length of the junction and $\xi$ the superconducting coherent length. In our W-Au NW-W junction, $\xi$ can be simply estimated by $\hbar v_F / 2\pi k_B T$ to be 300 nm at $T \sim 4$ K (here $v_F \sim 10^6$ m/s is the Fermi velocity) [4], thus, $\delta/\Delta \sim 0.2$ when the phase difference $\varphi$ between two superconducting electrodes is zero [7]. The minigap also depends on the phase difference $\varphi$ between the superconducting electrodes in the S-NW-S junction [10]. If we take $2\delta$ to be 92 µV, then $\delta/\Delta \sim 0.06$ is obtained for our W-Au NW-W junction and $\varphi$ between two superconducting electrodes is close to π. According to the experimental and analytical results of Fuechsle, *et al.*, the excitation of quasiparticles across the minigap is indeed more pronounced around phase difference $\varphi \sim$ π [42]. Interestingly, the ratio of $H_V/H_C \sim 0.05$ obtained in Ref. [4] is almost the same as the ratio of $\delta/\Delta$ ($H_V \sim 2.5$ kOe is the critical field to destroy the minigap and $H_C \sim 50$ kOe is the critical field to destroy the superconductivity of the electrodes at 2.6 K).

The Shapiro steps observed under microwave radiation in a "standard" Joshepson junction is temperature independent down to very low temperature [24-31]. In our experiment, the peaks in the "d$I$/d$V$" curves appear only between 2.1 and 3.7 K and most frequently between 2.4 and 3.6 K. The number of the peaks shows a maximum near 3 K, as shown in Fig. 2(c). The long-wavelength phonons, which has comparable energy with that of the minigap ~ 1.1 K, could be the candidate of excitation in our W-Au NW-W junction. Below 3 K, the number of peaks in the "d$I$/d$V$" curves decreases with decreasing the temperature, consisting with the idea that the acoustic phonons might not be sufficiently energetic to excite the quasiparticles across the minigap below 2 K since the long-wavelength phonons may be in lower energy than 1.1 K (the energy of the minigap). This model can also be applied to explain the disappearance of the resistance step at 2.5 kOe in the magneto-resistance measurements below 2.4 K, as reported in Ref. [4]. At temperature well above 3 K, a large fraction of the ABS immediately above $+\delta$ are occupied, thus, the excitation of quasiparticles across the minigap is suppressed and the number of peaks in the "d$I$/d$V$" curves decreases with increasing temperature. It appears the acoustic phonons are particularly effective in stimulating the quasi-particles across the minigap near 3 K. Very recently, Inoue, *et al.* demonstrated that superconducting quasiparticles in Josephson junction could strongly couple with long-wavelength phonons to form novel composite particles, the so-called Andreev polarons [43]. Their result further confirms that the long-wavelength phonons are the most reasonable candidate of excitation in our W-Au NW-W junction.

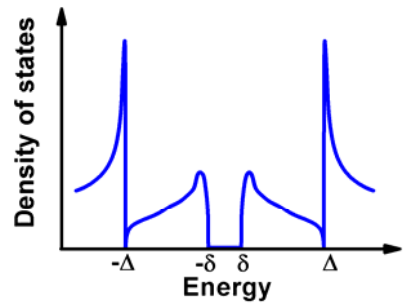

**FIG. 3.** Schematic density of states in our W-Au NW-W junction. The $\delta$ and $\Delta$ are defined in the text.

In summary, we observe Shapiro steps at voltages $V = (m/n)(2\delta/2e)$ in a S-NW-S Josephson junction in the absence of microwave radiation. This effect is attributed to the interplay between the ac Josephson current and the minigap of the Au nanowire. Our experimental results are consistent with the picture that the coupling between superconducting quasiparticles and the long-wavelength phonons in the Au nanowire is responsible for the appearance of the Shapiro steps. The results presented here also suggest the interesting possibility of detecting directly the microwave generated by the quantum jump of quasiparticles across the minigap.

This work is supported by National Natural Science Foundation of China (No. 11004010), the Fundamental Research Funds for the Central Universities, and the Penn State MRSEC under NSF grant DMR-0820404.

*Correspondence to: helin@bnu.edu.cn; jianwangphysics@pku.edu.cn; chan@phys.psu.edu.